\documentclass[preprint,prb,aps,floats,amssymb,showpacs,draft,%
floatfix,superscriptaddress]{revtex4} 
\usepackage{psfig}

\begin{document}

\title{ 
The 3D $\pm J$ Ising model at the ferromagnetic transition line
} 

\author{Martin Hasenbusch} 
\affiliation{ 
Dipartimento di Fisica dell'Universit\`a di Pisa and INFN, Pisa,
  Italy.  } 
\author{Francesco Parisen Toldin} 
\affiliation{ 
Scuola Normale Superiore and INFN, Pisa, Italy.  } 
\author{Andrea Pelissetto} 
\affiliation{Dipartimento di Fisica
  dell'Universit\`a di Roma ``La Sapienza" and INFN, Roma, Italy.}
\author{Ettore Vicari} 
\affiliation{ 
Dipartimento di Fisica dell'Universit\`a di Pisa and INFN, Pisa, Italy.  } 

\date{\today}

\begin{abstract}
  We study the critical behavior of the three-dimensional $\pm J$
  Ising model [with a random-exchange probability $P(J_{xy}) = p
  \delta(J_{xy} - J) + (1-p) \delta(J_{xy} + J)$] at the transition
  line between the paramagnetic and ferromagnetic phase, which extends
  from $p=1$ to a multicritical (Nishimori) point at $p=p_N\approx
  0.767$.  By a finite-size scaling analysis of Monte Carlo
  simulations at various values of $p$ in the region $p_N<p<1$, we
  provide strong numerical evidence that the critical behavior along
  the ferromagnetic transition line belongs to the same universality
  class as the three-dimensional randomly-dilute Ising
  model.  We obtain the results $\nu=0.682(3)$ and $\eta=0.036(2)$
  for the critical exponents, which are consistent with the estimates
  $\nu=0.683(2)$ and $\eta=0.036(1)$ at the transition of
  randomly-dilute Ising models.
\end{abstract}



\maketitle


\section{Introduction}
\label{intro}

The $\pm J$ Ising model has played an important role in the study of the
effects of quenched random disorder and frustration on Ising systems. It is
defined by the lattice Hamiltonian
\begin{equation}
{\cal H}_{\pm J} = - \sum_{\langle xy \rangle} J_{xy} \sigma_x \sigma_y,
\label{lH}
\end{equation}
where $\sigma_x=\pm 1$, the sum is over the nearest-neighbor sites of a simple
cubic lattice, and the exchange interactions $J_{xy}$ are uncorrelated
quenched random variables, taking values $\pm J$ with probability distribution
\begin{equation}
P(J_{xy}) = p \delta(J_{xy} - J) + (1-p) \delta(J_{xy} + J). 
\label{probdis}
\end{equation}
For $p=1$ we recover the standard Ising model, while for $p=1/2$ we obtain the
usual bimodal Ising spin-glass model.

\begin{figure*}[tb]
\centerline{\psfig{width=9truecm,angle=0,file=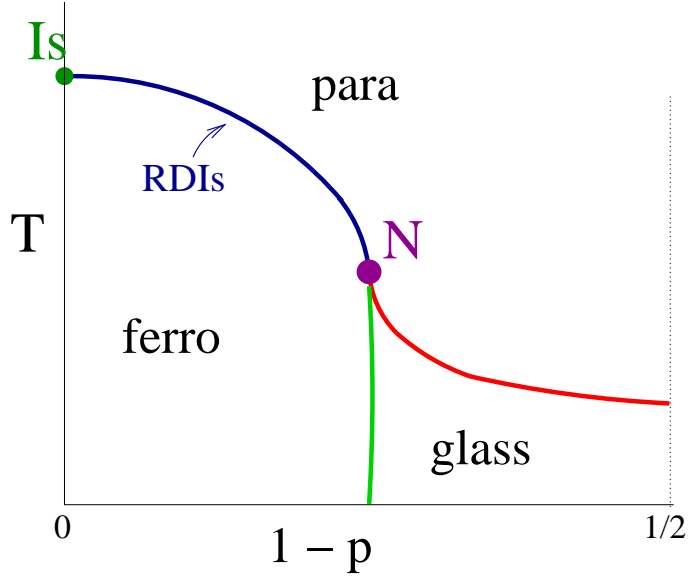}}
\vspace{2mm}
\caption{
Sketch of the phase diagram of the 3D $\pm J$ Ising model in the $T$-$p$
  plane.}
\label{phdia}
\end{figure*}

The phase diagram of the three-dimensional (3D) $\pm J$ Ising model is
sketched in Fig.~\ref{phdia}. The high-temperature phase is paramagnetic for
any $p$. The low-temperature phase depends on the value of $p$: it is
ferromagnetic for small values of $1-p$, while it is a spin-glass phase with
vanishing magnetization for sufficiently large values of $1-p$.  The different
phases are separated by transition lines, which meet at a multicritical point
$N$ located along the so-called Nishimori line.
\cite{Nishimori-81,LB-88,KR-03} The spin-glass transition has been mostly
studied at the symmetric point $p=1/2$, see, e.g.,
Refs.~\onlinecite{KR-03,KKY-06} and references therein. The spin-glass
transition line extends up to the Nishimori multicritical point,\cite{LB-88}
located at \cite{ON-87,Singh-91,OI-98,pnest} $p_N\approx 0.767$.  For larger
values of $p$, the transition is ferromagnetic, up to $p=1$ where one recovers
the pure Ising model, and therefore a transition in the Ising universality
class.  At the ferromagnetic transition line, for $p_N< p < 1$, the critical
behavior is expected to belong to a different universality class.

An interesting hypothesis, which has already been put forward in
Refs.~\onlinecite{Hukushima-00,KR-03}, is that the ferromagnetic transition of
the $\pm J$ Ising model belongs to the 3D randomly-dilute Ising (RDIs)
universality class (see, e.g., Refs.~\onlinecite{PV-02,FHY-03} for reviews on
randomly-dilute spin models).  A representative of the RDIs universality class
is the randomly site-dilute Ising model (RSIM) defined by the lattice
Hamiltonian
\begin{equation}
{\cal H}_{d} = - J\,\sum_{<xy>}  \rho_x \,\rho_y \; \sigma_x \sigma_y,
\label{Hdil}
\end{equation}
where $\rho_x$ are uncorrelated quenched random variables, which are
equal to $0,1$ with probability 
\begin{equation}
P(\rho_{x}) = p \delta(\rho_{x} - 1) + (1-p) \delta(\rho_{x}).
\end{equation}
For $p<1$ and above the percolation threshold of the spins ($p_{\rm
  perc}\approx 0.3116081(13)$ on a cubic lattice~\cite{BFMMPR-99}), the RSIM
undergoes a continuous phase transition between a disordered and a
ferromagnetic phase, whose nature is independent of $p$. This transition is
definitely different from the usual Ising transition: for instance, the
correlation-length critical exponent~\cite{HPPV-07,CMPV-03,PV-00,BFMMPR-98}
$\nu=0.683(2)$ differs from the Ising value~\cite{CPRV-02,DB-03}
$\nu=0.63012(16)$.  The RDIs universality class is expected to describe the
ferromagnetic transition in generic diluted ferromagnetic systems. For
instance, it has been verified that also the randomly bond-diluted Ising model
(RBIM) belongs to the RDIs universality class.\cite{HPPV-07,Janke-POS} These
results do not necessarily imply that also the $\pm J$ Ising model has an RDIs
ferromagnetic transition line. Indeed, while the RSIM (\ref{Hdil}) has only
ferromagnetic exchange interactions, the $\pm J$ Ising model is frustrated for
any value of $p<1$.  Therefore, the ferromagnetic transition in the $\pm J$
Ising model belongs to the RDIs universality class only if frustration is
irrelevant, a fact that is not obvious and should be carefully investigated.

Reference \onlinecite{Hukushima-00} investigated the issue by means of a Monte
Carlo (MC) renormalization-group (RG) study, claiming that the $\pm J$ Ising
model belongs to the same RDIs universality class as the RSIM and the RBIM. It
should be noted however that the quoted estimate for the correlation-length
exponent at the ferromagnetic transition, $\nu=0.658(9)$, is close to but not
fully consistent with the RDIs value $\nu=0.683(2)$.\cite{HPPV-07} Another
numerical MC work \cite{IOK-99} investigated the nonequilibrium relaxation
dynamics of the $\pm J$ Ising model and showed an apparent nonuniversal
dynamical critical behavior along the ferromagnetic transition line. These
results are not conclusive and further investigation is called for to clarify
this issue.
 
In this paper we focus on the transition line of the 3D $\pm J$ Ising model
between the paramagnetic and the ferromagnetic phase.  We investigate the
critical behavior by means of MC simulations at various values of $p$ in the
region $p_N<p<1$. Our finite-size scaling (FSS) analysis provides a strong
evidence that the critical behavior of the 3D $\pm J$ Ising along the
ferromagnetic line belongs to the 3D RDIs universality class.  For example, we
obtain $\nu=0.682(3)$ and $\eta=0.036(2)$, which are in good agreement with
the presently most accurate estimates \cite{HPPV-07} $\nu=0.683(2)$ and
$\eta=0.036(1)$ for the 3D RDIs universality class. 

The paper is organized as follows.  In Sec.~\ref{strategy} we summarize some
FSS results which are needed for the analysis of the MC data, and describe our
strategy to check whether the transition belongs to the RDIs universality
class. In Sec.~\ref{MC} we describe the MC simulations.  In Sec.~\ref{fss} we
report the results of the FSS analysis.  Finally, in Sec.~\ref{conclusions} we
draw our conclusions.  In App.~\ref{notations} we report the definitions of
the quantities we compute.

\section{Strategy of the finite-size scaling analysis}
\label{strategy}

In this work we check whether the ferromagnetic transition line in the
3D $\pm J$ Ising models belongs to the RDIs universality class.  For this
purpose, we present a FSS analysis of MC data for various values of $p$ in the
region $1>p >p_N\approx 0.767$.  We follow closely Ref.~\onlinecite{HPPV-07},
which studied the ferromagnetic transition line in the 3D RSIM and RBIM and
provided strong numerical evidence that these transitions belong to the same
RDIs universality class. We refer to Ref.~\onlinecite{HPPV-07} for notations
(a short summary is reported in App.~\ref{notations}) and a detailed
discussion of FSS in these disordered systems.

According to the RG, in the case of periodic boundary conditions and for $L\to
\infty$, where $L$ is the lattice size, a generic RG invariant quantity $R$ at
the critical temperature $1/\beta_c$ behaves as
\begin{equation}
R(L,\beta=\beta_c) = R^* \left( 1 + c_{11} L^{-\omega} + 
c_{12} L^{-2\omega} + \cdots + 
c_{21} L^{-\omega_2} + \cdots\right),
\label{Rscal}
\end{equation}
where $R^*$ is the universal infinite-volume limit and $\omega$ and $\omega_2$
are the leading and next-to-leading correction-to-scaling exponents.  In RDIs
systems scaling corrections play an important role,\cite{BFMMPR-98,CPPV-04}
since $\omega$ is quite small. Indeed we have $\omega=0.33(3)$ and
$\omega_2=0.82(8)$ in the 3D RDIs universality class.\cite{HPPV-07} These
slowly-decaying scaling corrections make the accurate determination of the
universal asymptotic behavior quite difficult.

Instead of computing the various quantities at fixed Hamiltonian parameters,
we keep a RG invariant quantity $R$ fixed at a given value
$R_{f}$.\cite{Has-99} This means that, for each $L$, we determine the
pseudocritical inverse temperature $\beta_f(L)$ such that
\begin{equation}
R(\beta=\beta_f(L),L) = R_{f}.
\label{rcbeta}
\end{equation}
All interesting thermodynamic quantities are then computed at $\beta =
\beta_f(L)$.  The pseudocritical inverse temperature $\beta_f(L)$ converges to
$\beta_c$ as $L\to \infty$.  The value $R_{f}$ can be specified at will, as
long as $R_f$ is taken between the high- and low-temperature fixed-point
values of $R$.  The choice $R_{f} = R^*$ (where $R^*$ is the critical-point
value) improves the convergence of $\beta_f$ to $\beta_c$ for $L\to\infty$;
indeed $\beta_f-\beta_c=O(L^{-1/\nu})$ for generic values of $R_f$, while
$\beta_f-\beta_c=O(L^{-1/\nu-\omega})$ for $R_f=R^*$.  This FSS method has
already been applied to the study of the critical behavior of $N$-vector spin
models, \cite{Has-99,CHPV-06} and of randomly-dilute Ising
models.\cite{HPPV-07}

As in Ref.~\onlinecite{HPPV-07}, we perform a FSS analysis at fixed
$R_\xi\equiv \xi/L =0.5943$, which is very close to the fixed-point value
$R_\xi^*=0.5944(7)$ of $R_\xi$ at $\beta_c$.  Given any RG invariant quantity
$R$, such as the quartic cumulants $U_4$ and $U_{22}$, we consider its value
at fixed $R_\xi$, i.e., $\bar{R}(L) = R(L,\beta_f(L))$. For $L\to \infty$,
$\bar{R}(L)$ behaves as $R(L,\beta_c)$:
\begin{equation}
{\bar R}(L) = {\bar R}^* \left( 1 
+ b_{11} L^{-\omega} + b_{12} L^{-2\omega} + \cdots + 
b_{21} L^{-\omega_2} + \cdots \right),
\label{Rscalbar}
\end{equation}
where the coefficients $b_{ij}$ depend on the Hamiltonian. 
The derivative ${\bar R}'$ with respect to $\beta$
of a generic RG invariant quantity $R$ behaves as
\begin{equation}
{\bar R}'(L) = a L^{1/\nu}
\left( 1 + a_{11} L^{-\omega} + a_{12} L^{-2\omega} + \cdots +
    a_{21} L^{-\omega_2} + \cdots\right) .
\label{Rprimeexp}
\end{equation}
Finally, the FSS of the magnetic susceptibility $\chi$ 
is given by~\cite{HPPV-07}
\begin{equation}
\bar{\chi}(L) \equiv  \chi(L,\beta = \beta_f(L)) = e L^{2-\eta} 
\left(  1 + e_{11} L^{-\omega} + e_{12} L^{-2\omega} + \cdots +
    e_{21} L^{-\omega_2} + \cdots\right) + e_b
\label{chiscal}
\end{equation}
where $e_b$ represents the background contribution.

A standard RG analysis, see, e.g., Ref.~\onlinecite{HPPV-07}, shows that the
amplitudes of the $O(L^{-k\omega})$ scaling corrections are proportional to
$u_3^k$ (with a universal coefficient), where $u_3$ is the leading irrelevant
scaling field with RG dimension $y_3=-\omega$.  Hamiltonians such that
$u_3=0$---we call them {\em improved} Hamiltonians---have a faster approach to
the universal asymptotic behavior, because the $O(L^{-k\omega})$ scaling
corrections vanish: $b_{1k} = a_{1k} = e_{1k} = 0$ in Eqs. (\ref{Rscalbar}),
(\ref{Rprimeexp}), and (\ref{chiscal}). In this case the leading scaling
corrections are proportional to $u_4 L^{-\omega_2}$, where $u_4$ is the
next-to-leading irrelevant scaling field and $y_4=-\omega_2$ is its RG
dimension. In Ref.~\onlinecite{HPPV-07} it was shown that the RSIM for $p =
p^* = 0.800(5)$ and the RBIM for $p = p^* = 0.56(2)$ are improved.  Since
scaling fields are analytic functions of the Hamiltonian parameters, $u_3$
must be proportional to $p-p^*$ close to $p=p^*$, i.e.  $u_3\approx c_3
(p-p^*)$.  Therefore, since the coefficients $b_{1k}$, $a_{1k}$, and $e_{1k}$
that appear in Eqs. (\ref{Rscalbar}), (\ref{Rprimeexp}), and (\ref{chiscal})
are proportional to $u_3^k$, we have
\begin{equation}
b_{1k},a_{1k},e_{1k}\sim (p-p^*)^k .
\label{arps}
\end{equation}
Beside the quantities defined in App.~\ref{notations}, 
we also consider observables---in analogy with the previous terminology,
we call them improved quantities---characterized by the fact
that the leading scaling correction proportional to $L^{-\omega}$
(approximately) vanishes in any model
belonging to the RDIs universality class.\cite{HPPV-07}
We consider the combination of quartic cumulants 
\begin{equation}
\bar{U}_{\rm im} = \bar{U}_{4}  + 1.3 \bar{U}_{22},
\label{uimdef}
\end{equation}
and improved estimators of the critical exponent $\nu$ defined as
\begin{equation}
R'_{\xi,{\rm im}} \equiv {\bar R}'_\xi \bar{U}_d^4, \qquad 
U'_{4,{\rm im}} \equiv {\bar U}'_4 \bar{U}_d^{2.5}
\label{impder}
\end{equation}
($\bar{U}_d$ is defined in App.~\ref{notations}).
In Ref.~\onlinecite{HPPV-07} we showed that, if the transition belongs to the
RDIs universality class, the leading scaling correction proportional to
$L^{-\omega}$ of these improved observables is suppressed.
More precisely, we showed that
the universal ratio of the amplitudes of the leading scaling correction in
$\bar{U}_{\rm im}$ and $\bar{U}_{4}$ satisfies
\begin{equation}
|{b_{11,\bar{U}_{\rm im}}/ b_{11,\bar{U}_{\rm 4}}}| \lesssim \frac{1}{15},
\label{b11r}
\end{equation}
while the one for the quantities $R'_{\xi,{\rm im}}$ and ${\bar R}'_\xi$ 
is bounded by  
\begin{equation}
|a_{11,R'_{\xi,{\rm im}}}/a_{11,{\bar R}'_\xi}| \lesssim \frac{1}{4}.
\label{e11r}
\end{equation}
The remaining scaling corrections are of order $L^{-2\omega}$ and
$L^{-\omega_2}$.  These improved observables are particular useful to check
whether the transition in a given system belongs to the 3D RDIs universality
class.

To summarize: in order to check whether the ferromagnetic transition of the 3D
$\pm J$ Ising model belongs to the RDIs universality class, we perform a FSS
analysis at fixed $R_\xi=0.5943$, and check if the results for the critical
exponents and other universal quantities are consistent with those obtained
for the RDIs universality class, which is characterized by \cite{HPPV-07}
critical exponents $\nu=0.683(2)$ and $\eta=0.036(1)$, by the leading and
next-to-leading scaling-correction exponents $\omega=0.33(3)$ and
$\omega_2=0.82(8)$ and by the universal infinite-volume values of the quartic
cumulants $\bar{U}_{22}^*=0.148(1)$, $\bar{U}_{\rm im}^*=1.840(4)$, and
$\bar{U}_{d}^*=1.500(1)$.  Notice that the fact that we fix $R_\xi=0.5943$
does not introduce any bias in our FSS analysis.

\section{Monte Carlo simulations}
\label{MC}

\begin{figure*}[tb]
\centerline{\psfig{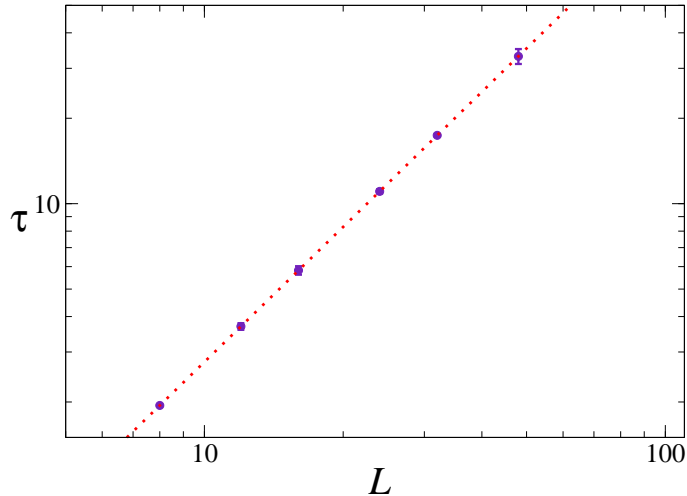}}
\vspace{2mm}
\caption{
  Exponential autocorrelation time $\tau$ of the magnetic susceptibility for a
  mixture of Metropolis and cluster updates as discussed in the text, at
  $p=0.87$.  The dotted line shows the result of a fit to $\tau= c L^z$: this
  fit gives $z\approx 1.6$.  }
\label{taucluster}
\end{figure*}

We performed MC simulations of Hamiltonian (\ref{lH}) with $J=1$ for
$p=0.94,0.90,0.883,0.87,0.83,0.80$, close to the critical temperature on cubic
lattices of size $L^3$ with periodic boundary conditions, for a large range of
lattice sizes: from $L=8$ to $L=80$ for $p=0.883,0.87$, to $L=64$ for
$p=0.94,0.90,0.83$, and to $L=48$ for $p=0.80$.  We chose values of $p$ not
too close to $p=1$: indeed, as $p\to 1$ we expect crossover effects due to the
presence of the Ising transition for $p=1$ and, therefore, that the asymptotic
behavior sets in only for large values of $L$. We return to this point 
later.

We used a Metropolis algorithm and multispin coding.  \cite{multispin} In the
simulation $n_{\rm bit}$ systems evolve in parallel, where $n_{\rm bit}=32$ or
$n_{\rm bit}=64$ depending on the computer that is used. For each of these
$n_{\rm bit}$ systems we use a different set of couplings $J_{xy}$.  This
allows us to perform 64 parallel simulations on a 64-bit machine, and
therefore to gain a large factor in the efficiency of the MC simulations.  We
used high-quality random-number generators, such as the RANLUX \cite{ranlxd}
or the twister \cite{twister} generators.~\cite{footnote1} Using the twister
random-number generator, we need about $1.2 \times 10^{-9}$ seconds for one
Metropolis update of a single spin on an Opteron processor running at 2 GHz.
Our simulations took approximately 3 CPU years on an Opteron (2 GHz)
processor.

It is worth mentioning that cluster algorithms, such as the Swendsen-Wang
cluster \cite{SW-87} and the Wolff single-cluster \cite{Wolff-89} algorithm,
show significant slowing down in the $\pm J$ Ising model. At the earlier stage
of this work we performed some simulations of the $\pm J$ Ising model at
$p=0.87$ using the algorithm used in Ref.~\onlinecite{HPPV-07} to simulate the
RSIM and the RBIM.  There we used a combination of Metropolis, Swendsen-Wang
cluster,\cite{SW-87} and Wolff single-cluster \cite{Wolff-89} updates.  More
precisely, each updating step consisted of 1 Swendsen-Wang update, 1
Metropolis update, and $L$ single-cluster updates.  In all cases the
exponential autocorrelation times $\tau$ of the magnetic susceptibility was
small: $\tau\lesssim 1$ in units of the above updating step, even for the
largest lattice sizes considered, i.e.  $L=192$.  In the $\pm J$ Ising model
at $p=0.87$ autocorrelation times are much larger.  In Fig.~\ref{taucluster}
we plot estimates of $\tau$ as obtained from the magnetic susceptibility. They
show a clear evidence of critical slowing down: $\tau \sim L^z$ with $z\approx
1.6$.  Such a value of $z$ should be compared with the dynamic exponent of
Swendsen-Wang and Wolff cluster algorithms in the RSIM, which is much smaller:
\cite{IIBH-06} $z\lesssim 0.5$.  These results show that cluster algorithms
behave differently in the $\pm J$ Ising model, likely due to frustration.
They suggest that frustration is relevant for the cluster dynamics.

Taking also into account the computer time required by the cluster algorithms,
we then turned to a multispin Metropolis algorithm. This turns out to be much
more effective at the lattice sizes considered, although it has a larger
dynamic exponent $z\gtrsim 2$, see, e.g., Ref.~\onlinecite{IIBH-06} and
references therein.  We also mention that the autocorrelation time
significantly increases with decreasing $p$ (keeping $L$ fixed).  For example,
for $L=48$ it increases by approximately a factor of 10 from $p=0.90$ to
$p=0.80$.  This represents a major limitation to perform simulations for large
lattices close to the multicritical point.

For each lattice size we considered $N_s$ disorder samples, with $N_s$
decreasing with increasing $L$, from $N_s \gtrsim 10^6$ for $L=8$ to $N_s
\gtrsim 2\times 10^4$ for the largest lattices.  For each disorder sample, we
collected a few hundred independent measurements at equilibrium.  The averages
over disorder are affected by a bias due to the finite number of measures at
fixed disorder.\cite{BFMMPR-98-b,HPPV-07} A bias correction is required
whenever one considers the disorder average of combinations of thermal
averages.  We used the formulas reported in App. B of
Ref.~\onlinecite{HPPV-07}.  Errors were computed from the sample-to-sample
fluctuations and were determined by using the jackknife method.~\cite{footnote1} 

Our FSS analysis is performed at fixed $R_\xi\equiv \xi/L$.  In order to
determine expectation values at fixed $R_\xi$, one needs the values of the
observables as a function of $\beta$ in some neighborhood of the inverse
temperature $\beta_{\rm run}$ used in the simulation.  In
Ref.~\onlinecite{HPPV-07} we used the reweighting method for this purpose.
This requires that the observables and, in particular, the values of the
energy are stored at each measurement. For the huge statistics like those we
have for the smaller values of $L$, this becomes unpractical.  Therefore, we
used here a second-order Taylor expansion, determining $O(\beta,L)$ from
$O(\beta_{\rm run},L) + a_O (\beta - \beta_{\rm run}) + b_O (\beta -
\beta_{\rm run})^2$.  The coefficients $a_O$ and $b_O$ are obtained from
appropriate expectation values as in Ref.~\onlinecite{CHPV-06}.  Since their
computation involves disorder averages of products of thermal averages, we
have implemented in all cases an exact bias correction, using the formulas of
Ref.~\onlinecite{HPPV-07}.  Derivatives with respect to $\beta$ are then
obtained as $O'(\beta,L) = a_O + 2 b_O (\beta - \beta_{\rm run})$.  Of course,
this method requires $|\beta_{\rm run} - \beta_f|$ to be sufficiently small.
We have carefully checked the results by performing, for each $L$ and $p$,
runs at different values of $\beta$.

The MC estimates of the quantities introduced in Sec.~\ref{strategy} and in
App.~\ref{notations} at fixed $R_\xi\equiv \xi/L = 0.5943$ are available on
request.

\section{Finite-size scaling analysis}
\label{fss}

In this section we present the results of our FSS analysis of the MC data at
fixed $R_\xi=0.5943$.

\subsection{Renormalization-group invariant quantities}
\label{rginv}

\begin{figure*}[tb]
\centerline{\psfig{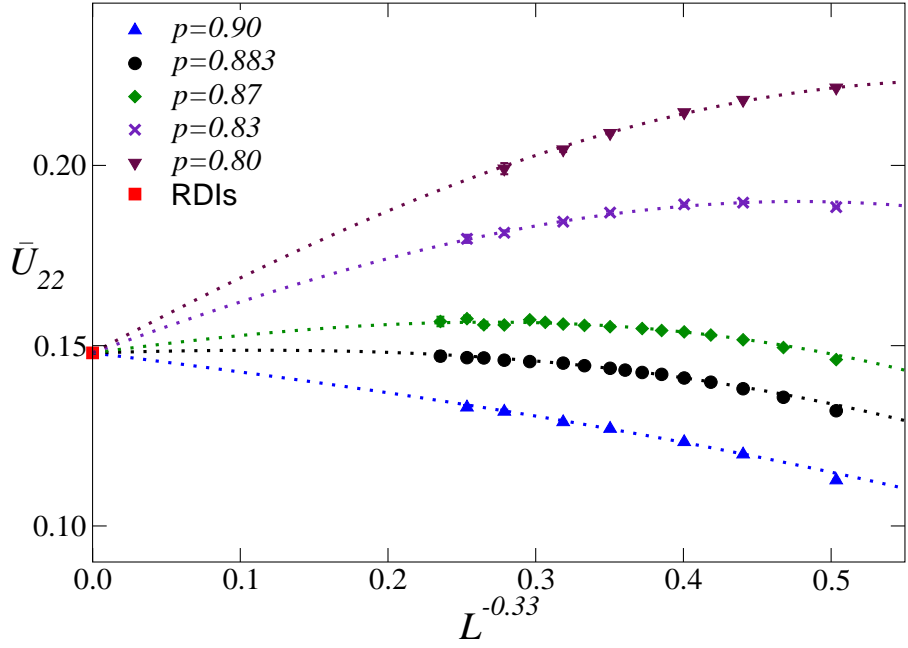}}
\vspace{2mm}
\caption{
  MC estimates of ${\bar U}_{22}$ versus $L^{-\omega}$ with $\omega=0.33$
  for different values of $p$.  The dotted
  lines show results of fits to $c_0 + c_1 L^{-\varepsilon_1} + c_2
  L^{-\varepsilon_2}$, fixing $c_0=0.148$, $\varepsilon_1=0.33$, and
  $\varepsilon_2=0.82$. In the RDIs universality class ${\bar U}_{22}^* = 0.148(1)$.
  \protect\cite{HPPV-07} }
\label{u22fig}
\end{figure*}

In Fig.~\ref{u22fig} we show the MC estimates of $\bar{U}_{22}$ versus
$L^{-\omega}$ with $\omega=0.33(3)$, which is the leading scaling exponent of
the RDIs universality class. The data vary significantly with $p$ and $L$.
This $p$ and $L$ dependence is always consistent with the existence of the
expected next-to-leading scaling corrections, i.e. with a behavior of the form
\begin{equation}
\bar{U}_{22} = \bar{U}_{22}^* + 
c_1 L^{-\varepsilon_1} + c_2 L^{-\varepsilon_2},
\label{f5ans}
\end{equation}
where $\bar{U}_{22}^*$, $\varepsilon_1$ and $\varepsilon_2$ are fixed to the
RDIs values:\cite{HPPV-07} $\bar{U}_{22}^*=0.148$, $\varepsilon_1=0.33$ and
$\varepsilon_2=0.66,0.82$. The fits corresponding to $\varepsilon_2 = 0.82$
are shown in Fig.~\ref{u22fig}. Note that in most of the cases it is crucial
to include a next-to-leading correction.  Only for $p=0.90$  the data are well
fitted by taking only the leading scaling correction.

An unbiased estimate of $\omega$ can be obtained from the difference
of data at different values of $p$, i.e. by considering
\begin{equation}
\bar{U}_{22}(p_1;L) - \bar{U}_{22}(p_2;L) \approx c L^{-\omega}.
\label{diffo}
\end{equation}
Linear fits of the logarithm of these differences give results in reasonable
agreement with the RDIs estimate $\omega=0.33(3)$, especially when only data
corresponding to $L\ge L_{\rm min}=24$ are used.  For $L_{\rm min}=24$
[$L_{\rm min}=32$], we obtain $\omega=0.27(2)$ [$\omega=0.27(3)$] from the
data at $p_1=0.83$ and $p_2=0.90$, $\omega=0.19(5)$ [$\omega=0.31(9)$] from
those at $p_1=0.883$ and $p_2=0.90$, and $\omega=0.29(3)$ [$\omega=0.25(4)$]
from the results at $p_1=0.83$ and $p_2=0.883$.  We also fitted the difference
$\bar{U}_{22}-0.148$ at $p=0.90$ to $c L^{-\varepsilon}$ (for this value of
$p$ next-to-leading corrections are apparently very small, see
Fig.~\ref{u22fig}).  We obtain $\omega=0.35(3)$ [$\omega=0.39(6)$] for
$L_{\rm min}=24$ [$L_{\rm min}=32$].

\begin{figure*}[tb]
\centerline{\psfig{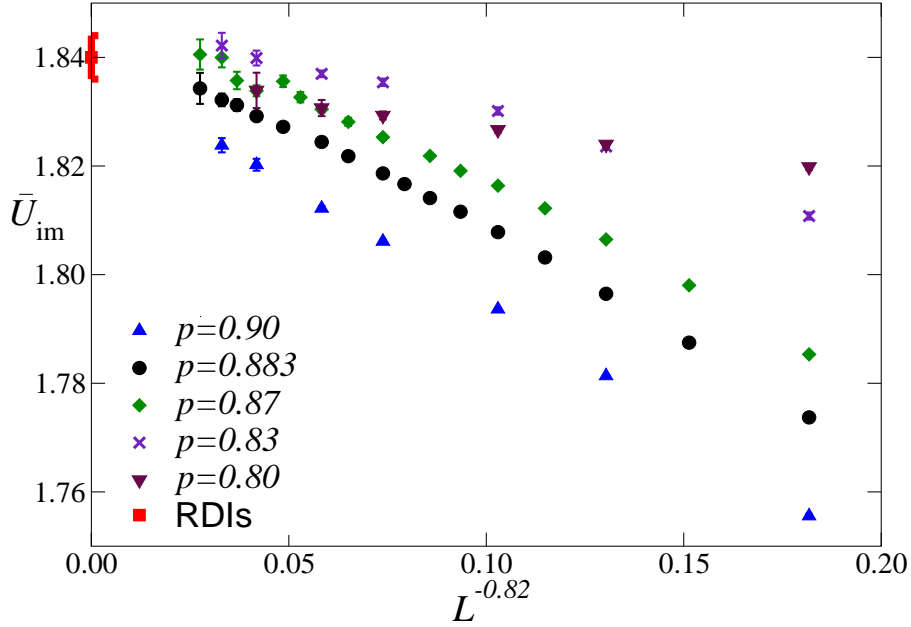}}
\vspace{2mm}
\caption{
MC estimates of ${\bar U}_{\rm im}$ versus $L^{-0.82}$.  
The filled square on the vertical axis corresponds to 
the RDIs estimate~\cite{HPPV-07} ${\bar U}_{\rm im}^*=1.840(4)$.
}
\label{uimfig}
\end{figure*}

\begin{figure*}[tb]
\centerline{\psfig{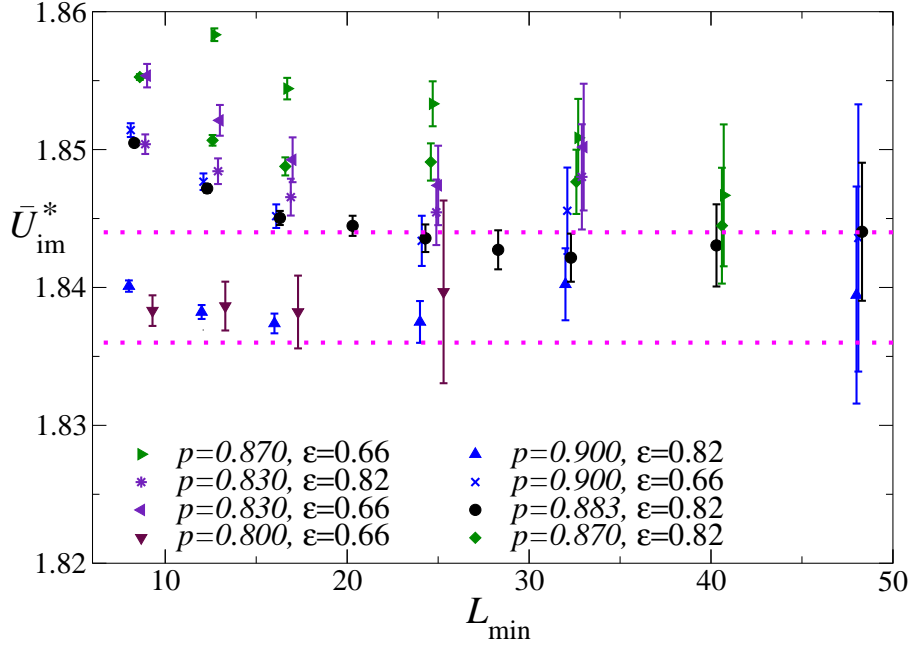}}
\vspace{2mm}
\caption{
Estimates of ${\bar U}_{\rm im}^*$ as obtained by
fits to ${\bar U}_{\rm im}^*+c
L^{-\varepsilon}$, versus  the minimum lattice size $L_{min}$ allowed in the fits.
Some data are slightly shifted along the $x$-axis to make them visible.
The dotted lines correspond to the RDIs estimate \cite{HPPV-07} $\bar{U}_{\rm
im}^*=1.840(4)$.
}
\label{uimfitfig}
\end{figure*}

The results of the above-reported fits of $\bar{U}_{22}$ show that the leading
scaling corrections proportional to $L^{-\omega}$ vanish for $p\approx 0.883$.
Note that, close to $p^*$, the relevant next-to-leading scaling corrections
should be those proportional to $L^{-\omega_2}$ with $\omega_2\approx 0.82$.
Indeed, according to Eq.~(\ref{arps}), the coefficient of those proportional
to $L^{-2\omega}$ is of order $(p-p^*)^2$, i.e. $b_{12}\approx \bar{b}_{12}
(p-p^*)$. Therefore, $b_{12}$ is small if $\bar{b}_{12}=O(1)$ (we checked this
numerically). This applies to the FSS at $p=0.87$ and 0.90, where the
$L^{-2\omega}$ corrections can be neglected, although in these two cases we
cannot neglect the leading $L^{-\omega}$ correction whose coefficient is
proportional to $p-p^*$.  An analysis of the leading scaling corrections at
$p=0.87,0.883,0.90$, assuming the RDIs values $\bar{U}_{22}^*=0.148(1)$ and
$\omega=0.33(3)$ (we perform combined fits to (\ref{f5ans}) with
$\varepsilon_1 = \omega$) gives the estimate
\begin{equation}
p^*=0.883(3),
\label{pstar}
\end{equation}
which is approximately in the middle of the ferromagnetic line,
i.e. $1-p^*\approx (1-p_N)/2$. We performed a similar 
analysis for $\bar{U}_d$, obtaining a consistent estimate of $p^*$.
Thus, the $\pm J$ Ising model for $p=0.883$ is approximately
improved.  Therefore, at $p=0.883$, fits of the data assuming
$O(L^{-\omega_2})$ leading scaling corrections should provide reliable
results.

As discussed in Sec.~\ref{strategy}, a useful quantity to perform stringent
checks of universality within the RDIs universality class is the combination
${\bar U}_{\rm im}$ of quartic cumulants reported in Eq.~(\ref{uimdef}). For
this quantity the scaling corrections proportional to $L^{-\omega}$ are small,
cf. Eq.~(\ref{b11r}), and thus the dominant corrections should behave as
$L^{-2\omega}$, with $2\omega\approx 0.66$.  As already discussed, for values
of $p$ close to $p^*$, such as $p=0.87,0.883,0.90$, also the $L^{-2\omega}$
term is expected to be small and thus the dominant corrections should scale as
$L^{-\omega_2}$ with $\omega_2\approx 0.82$.  In Fig.~\ref{uimfig} we show the
MC results for ${\bar U}_{\rm im}$ for various values of $p$.
Fig.~\ref{uimfitfig} shows results of fits to
\begin{equation}
\bar{U}^* + c L^{-\varepsilon},
\label{f4ans}
\end{equation}
with $\varepsilon=0.66,0.82$.  We obtain $\bar{U}_{\rm im}^*= 1.840(3)[3],\;
1.842(2)[1],\; 1.845(2)[3]$ respectively for $p=0.90,\;0.883,\;0.87$, fixing
$\varepsilon=\omega_2=0.82(8)$ (the error in brackets is related to the
uncertainty of $\omega_2$) and using data with $L\ge 32$; moreover we obtain
$\bar{U}_{\rm im}^*= 1.847(3)[2],\; 1.840(7)[1]$ respectively for
$p=0.83,\;0.80$, fixing $\varepsilon=0.66(6)$ and using data with $L\ge 24$.
For all values of $p$ the results are in good agreement with RDIs estimate
\cite{HPPV-07} $\bar{U}_{\rm im}^*=1.840(4)$.  They provide strong support to
a RDIs critical behavior along the ferromagnetic line.

\begin{figure*}[tb]
\centerline{\psfig{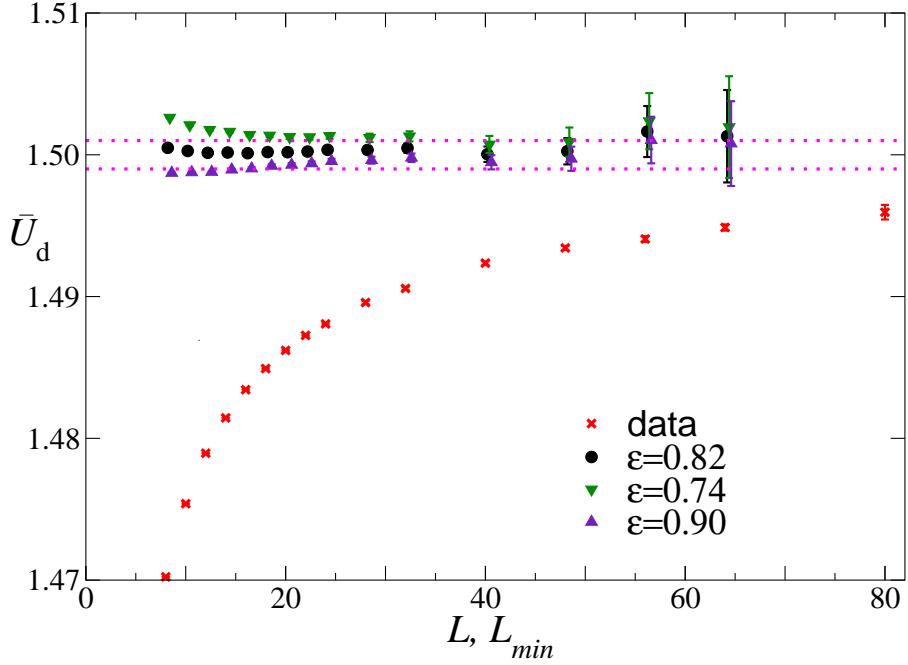}}
\vspace{2mm}
\caption{
  Estimates of ${\bar U}_{d}^*$ as obtained by fits of ${\bar U}_{d}$ at
  $p=0.883$, to ${\bar U}_{\rm d}^*+c L^{-\varepsilon}$.  Some data are
  slightly shifted along the $x$-axis to make them visible.  The dotted lines
  correspond to the RDIs estimate~\cite{HPPV-07} $\bar{U}_{d}^*=1.500(1)$.  }
\label{udfig}
\end{figure*}

A further stringent check of universality comes from the analysis of the data
$\bar{U}_d$ at $p=0.883$, because the data of $\bar{U}_d$ are very precise due
to a cancellation of the statistical fluctuations.\cite{HPPV-07} Since the
model is improved, the $L^{-k\omega}$ scaling corrections are negligible and
the large-$L$ behavior is approached with corrections of order
$L^{-\omega_2}$, $\omega_2 = 0.82(8)$. We thus fit the data to
Eq.~(\ref{f4ans}) with $\varepsilon=0.74,0.82,0.90$.  In Fig.~\ref{udfig} we
show the results. We obtain $\bar{U}_d^*=1.5001(1)[15],\;
1.5004(2)[9],\;1.5003(10)[6]$ (the error in brackets is related to the
uncertainty of $\omega_2=0.82(8)$) for $L_{\rm min}=12,\;24,\;48$
respectively.  Moreover, by fitting the data to $\bar{U}_d^* + c_1
L^{-\varepsilon_1} + c_2 L^{-\varepsilon_2}$ with $\varepsilon_1=0.33$ and
$\varepsilon_2=0.82$, we obtain $\bar{U}_d^*=1.5006(7),\; 1.500(3)$ for
$L_{\rm min}=12,\;24$ respectively.  These results are in perfect agreement
with the RDIs estimate \cite{HPPV-07} $\bar{U}_{d}^*=1.500(1)$.  Such an
agreement is also confirmed by the analysis of the data of $\bar{U}_{22}$, for
example a fit to Eq.~(\ref{f4ans}) with $\varepsilon=\omega_2=0.82(8)$ gives
$\bar{U}_{22}^*=0.1486(8)[3]$ for $L_{\rm min}=32$,
to be compared with the RDIs estimate~\cite{HPPV-07} 
$\bar{U}_{22}^*=0.148(1)$.

We have not shown results for values of $p$ too close to 1, for $p>0.90$ say,
because they are affected by crossover effects due to presence of the Ising
transition for $p=1$, as it also occurs in randomly dilute Ising models.
\cite{BFMMPR-98,CPPV-04,footnotecrossover} For instance, for $p=0.94$ the data
are not compatible with a behavior of the form (\ref{f5ans}) with
$\bar{U}_{22}^*$ fixed to the RDIs value. Our data that correspond to lattice
sizes $L\le 64$ apparently converge to a smaller value, consistently with the
expected crossover from pure to random behavior (in pure systems
$\bar{U}_{22}^*=0)$.  The same quantitative differences are observed in the
RSIM and in the RBIM close to the Ising transition.  This suggests that in FSS
analyses up to $L\approx 100$ the asymptotic RDIs behavior can only be
observed for $p \lesssim 0.94$.

\subsection{Critical exponents}
\label{critexp}

\begin{figure*}[tb]
\centerline{\psfig{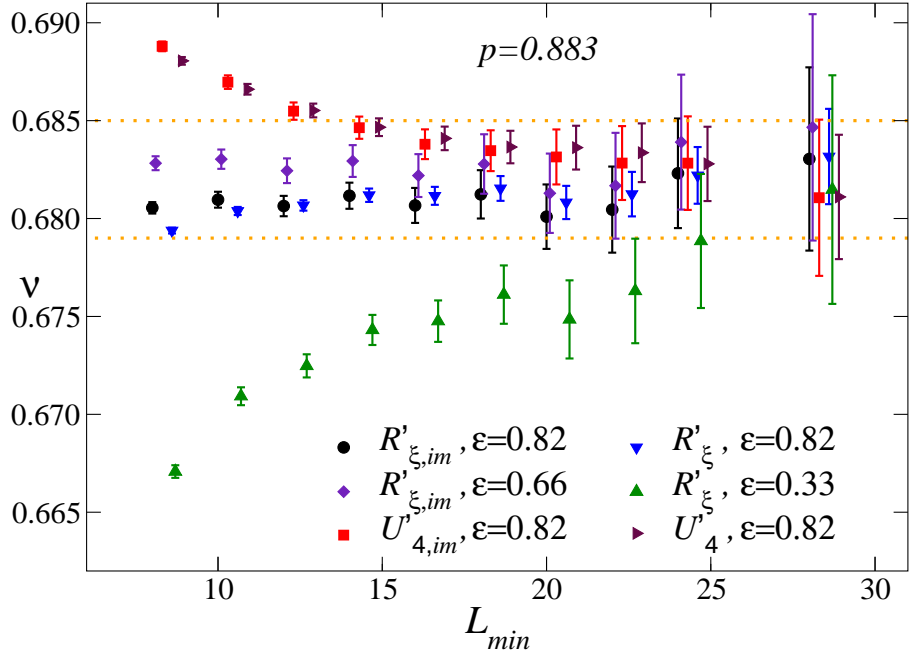}}
\vspace{2mm}
\caption{
  Estimates of the critical exponent $\nu$, as obtained by fits of 
  ${\bar R}'_\xi$, ${\bar U}'_{4}$, $R'_{\xi,{\rm im}}$, and
  $U'_{4,{\rm im}}$ at $p=0.883$.  
  $L_{\rm min}$ is the minimum lattice size allowed in the fits.
Some data are slightly shifted along the $x$-axis to make them visible.
The dotted lines correspond to the estimate $\nu=0.682(3)$.  }
\label{nu0.883fig}
\end{figure*}

The correlation-length exponent $\nu$ can be estimated by fitting the
derivative of $R_\xi$ and $U_4$ to the expression (\ref{Rprimeexp}).  Accurate
estimates are only obtained for improved Hamiltonians.  For generic models, as
shown in Ref.~\onlinecite{HPPV-07}, good estimates are only obtained by using
improved estimators, such as those reported in Eq.~(\ref{impder}).

We analyze the data at $p=0.883$, which is a very good approximation of the
improved value $p^*=0.883(3)$.  In Fig.~\ref{nu0.883fig} we report several
results for the critical exponent $\nu$, obtained by analyzing ${\bar
  R}'_\xi$, ${\bar U}'_{4}$, and their improved versions $R'_{\xi,{\rm im}}$
and $U'_{4,{\rm im}}$.  We show results of fits of their logarithms to
\begin{equation}
{1\over \nu} {\rm ln} L + a + b L^{-\varepsilon},
\label{f1ans}
\end{equation}
fixing $\varepsilon$ to several values.  Since the Hamiltonian is
approximately improved, scaling corrections are expected to decrease as
$L^{-\omega_2}$ with $\omega_2=0.82(8)$.  Since $p=0.883$ is only
approximately equal to $p^*$, one may be worried of the residual leading
scaling corrections that are small but do not vanish exactly.  Improved
estimators should provide the most reliable results since the leading scaling
corrections are additionally suppressed.

As can be seen in Fig.~\ref{nu0.883fig}, the results obtained by using ${\bar
  R}'_\xi$ and ${\bar U}'_{4}$ and $\varepsilon = 0.82$ are perfectly
consistent with those obtained from their improved versions. This confirms
that the Hamiltonian is improved.  Fits of ${\bar R}'_\xi$ to (\ref{f1ans})
with $\varepsilon=0.33$ do not provide stable results.  The results approach
the values obtained in the other fits only when increasing the minimum size
$L_{\rm min}$ allowed in the fit. This is expected, since the $L^{-\omega}$
corrections should be negligible with respect to the $L^{-\omega_2}$ ones. In
conclusion, our final estimate of the correlation-length exponent is
\begin{equation}
\nu=0.682(3),
\label{nuest}
\end{equation}
which includes all results (with their errors) of the fits of ${\bar R}'_\xi$,
${\bar U}'_{4}$, $R'_{\xi,{\rm im}}$, $U'_{4,{\rm im}}$ to Eq.~(\ref{f1ans})
with $\varepsilon=0.82(8)$ and $L_{\rm min} =16,24$.  Estimate (\ref{nuest})
is in perfect agreement with the most precise RDIs estimate $\nu=0.683(2)$.

\begin{figure*}[tb]
\centerline{\psfig{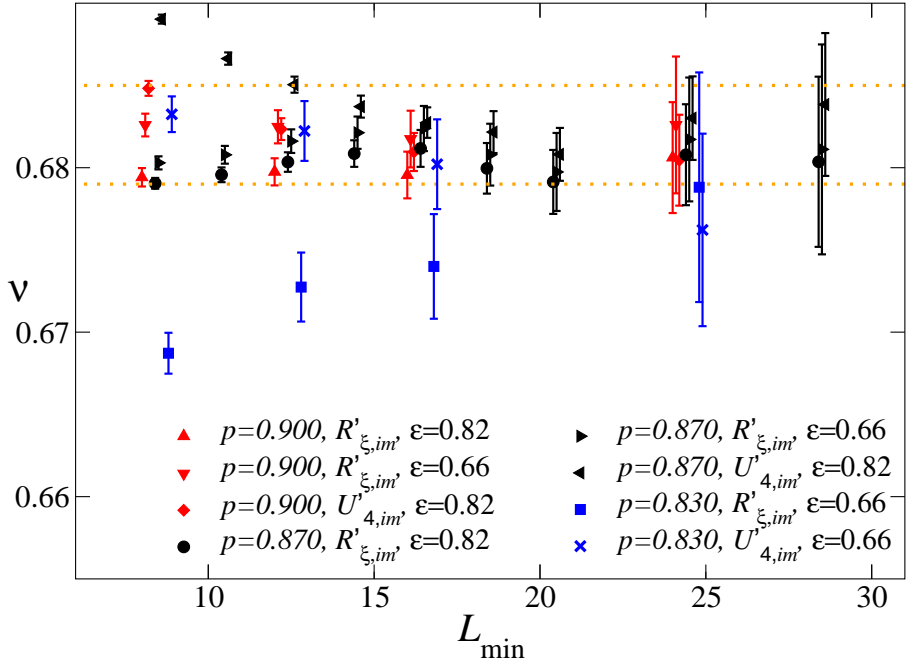}}
\vspace{2mm}
\caption{
  Results from $\nu$ obtained by fitting $R'_{\xi,{\rm im}}$ to $a L^{1/\nu}
  (1 + b L^{-\varepsilon})$.  Some data are slightly shifted along the
  $x$-axis to make them visible.  The dotted lines correspond to the estimate
  obtained at $p=0.883$, i.e. $\nu=0.682(3)$.  }
\label{nuallfig}
\end{figure*}

Estimate (\ref{nuest}) is also confirmed by the analysis of the data at the
other values of $p$.  Fig.~\ref{nuallfig} shows results obtained by fitting
the logarithm of $R'_{\xi,{\rm im}}$ to the function (\ref{f1ans}) for other
values of $p$.  They are definitely consistent with the result obtained at
$p=0.883$.  Results for $p=0.80$ are not shown because the available data are
not sufficient to get reliable results.

\begin{figure*}[tb]
\centerline{\psfig{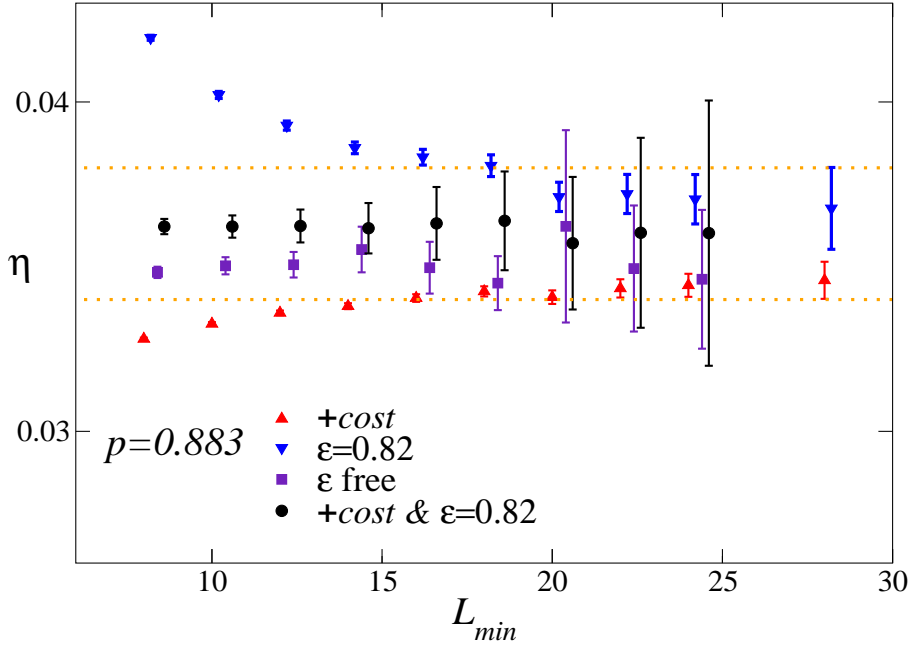}}
\vspace{2mm}
\caption{
  Estimates of the critical exponent $\eta$, obtained by fitting $\ln\bar\chi$
  at $p=0.883$, to $a + (2-\eta) \ln L + b L^{\eta-2}$ (denoted by +$cost$),
  $a + (2-\eta) \ln L + a_1 L^{-\varepsilon}$, and to $a + (2-\eta) \ln L +
  a_1 L^{-\varepsilon} + b L^{\eta-2}$.  Some data are slightly shifted along
  the $x$-axis to make them visible.  The dotted lines correspond to the final
  estimate $\eta=0.036(2)$.  }
\label{eta0.883fig}
\end{figure*}

\begin{figure*}[tb]
\centerline{\psfig{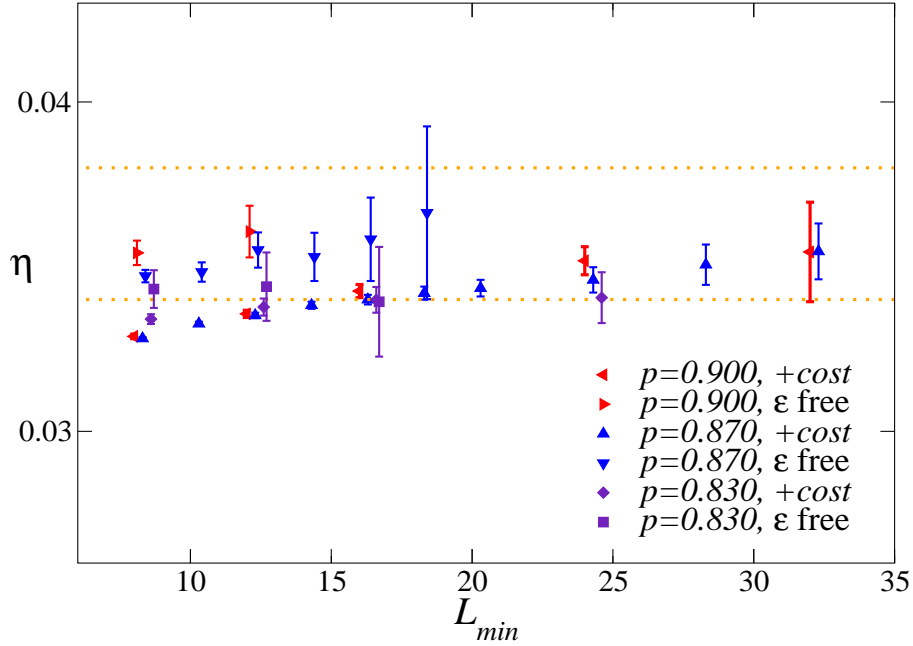}}
\vspace{2mm}
\caption{
  Estimates of the critical exponent $\eta$, obtained by fitting $\bar\chi$
  for various values of $p$.  See the caption of
  Fig.~\protect\ref{eta0.883fig} for an explanation of the fits.  Some data
  are slightly shifted along the $x$-axis to make them visible.  The dotted
  lines correspond to the result $\eta=0.036(2)$ obtained at $p=0.883$.  }
\label{etaothersfig}
\end{figure*}

In order to estimate the critical exponent $\eta$, we analyze the FSS of the
magnetic susceptibility $\bar{\chi}$, cf. Eq.~(\ref{chiscal}).  We fit it to
$a L^{2-\eta} + b $ (where $b$ represents a constant background term), to $a
L^{2-\eta} (1 + c L^{-\varepsilon})$, and to $a L^{2-\eta} (1 + c
L^{-\varepsilon})+ b$ (more precisely, we fit $\ln\bar\chi$ to the logarithm
of the previous expressions).  The results at $p=0.883$ are shown in
Fig.~\ref{eta0.883fig}, versus the minimum size $L_{\rm min}$ allowed in the
fits. We obtain the estimate
\begin{equation}
\eta=0.036(2),
\label{etaest}
\end{equation}
which includes all results obtained for $L_{\rm min}\gtrsim 16$.  This
estimate agrees with the most precise RDIs estimate $\eta=0.036(1)$.
Fig.~\ref{etaothersfig} shows results for the other values of $p$.  Again,
they are in good agreement.

\subsection{The critical temperature}
\label{crittc}

\begin{figure*}[tb]
\centerline{\psfig{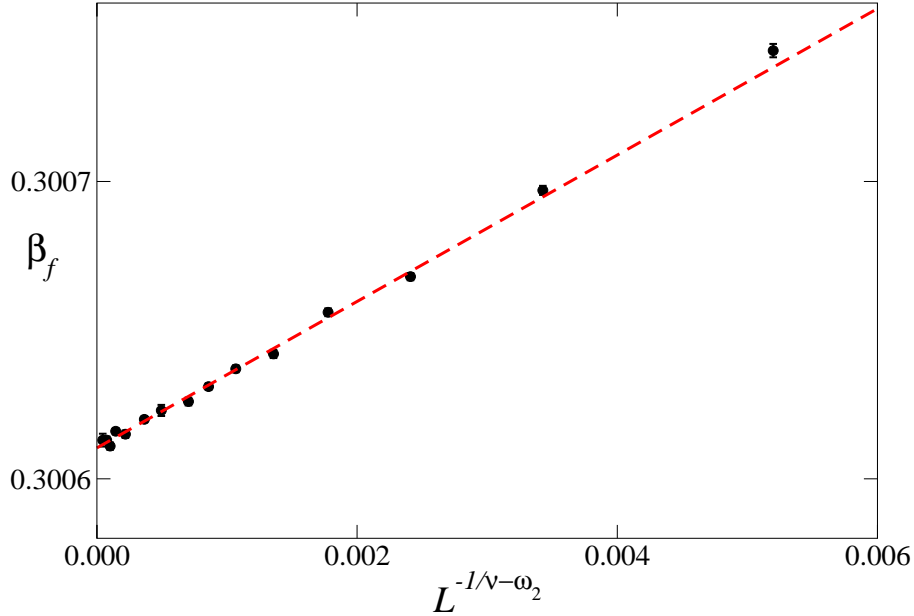}}
\vspace{2mm}
\caption{
  Estimates of $\beta_f(L)$ at $p=0.883$ versus $L^{-(1/\nu+\omega_2)}$ for
  $1/\nu + \omega_2\approx 2.28$.  The dashed line corresponds to a linear fit
  of the data for $L\ge 12$.  }
\label{bf0.883fig}
\end{figure*}

The critical temperature can be estimated by extrapolating the estimates of
$\beta_f$ at $R_\xi=0.5943$, cf.~Eq.~(\ref{rcbeta}).  Since we have chosen
$R_\xi=0.5943\approx R_\xi^*=0.5944(7)$,~\cite{HPPV-07} we expect in general that
$\beta_f-\beta_c=O(L^{-1/\nu-\omega})$.  For $p=0.883$, since the model is
approximately improved, the leading scaling corrections are related to the
next-to-leading exponent $\omega_2$.  Thus, in this case
$\beta_f-\beta_c=O(L^{-1/\nu-\omega_2})$. This behavior is nicely observed in
Fig.~\ref{bf0.883fig}, which shows $\beta_f(L)$ at $p=0.883$ vs
$L^{-1/\nu-\omega_2}$ with $1/\nu+\omega_2\approx 2.28$. A fit to $\beta_c + a
L^{-1/\nu-\omega_2}$ gives $\beta_c=0.300611(1)$.

\begin{figure*}[tb]
\centerline{\psfig{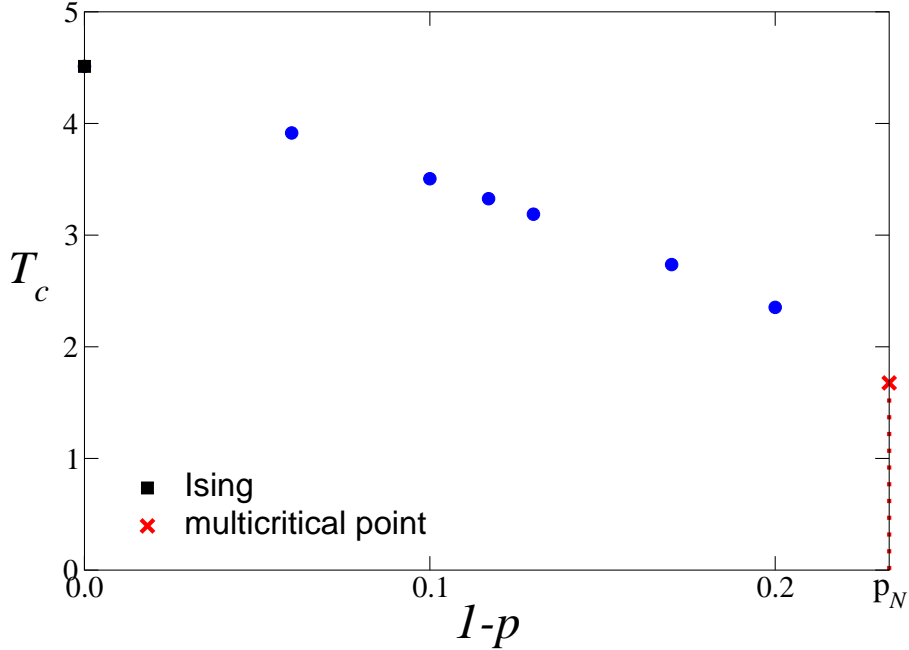}}
\vspace{2mm}
\caption{
The critical temperature $T_c\equiv 1/\beta_c$ vs $1-p$.
}
\label{bcfig}
\end{figure*}

For the other values of $p$ we expect $\beta_f-\beta_c=O(L^{-1/\nu-\omega})$
with $1/\nu+\omega\approx 1.79$. Linear fits of $\beta_f(L)$ (for $L\ge L_{\rm
  min}$ with $L_{\rm min}$ sufficiently large to give an acceptable $\chi^2$)
give the estimates $\beta_c=0.25544(2)$ for $p=0.94$, $\beta_c=0.285285(5)$
for $p=0.90$, $\beta_c=0.313748(1)$ for $p=0.87$, $\beta_c=0.365459(5)$ for
$p=0.83$, $\beta_c=0.42501(3)$ for $p=0.80$.  We finally recall that
\cite{DB-03} $\beta_c=0.22165452(8)$ for $p=1$ (the standard Ising model), and
that \cite{OI-98} $\beta_c=0.5967(11)$ at the multicritical Nishimori point at
$p_N=0.7673(3)$.  In Fig.~\ref{bcfig} we plot the available estimates of the
critical temperature $T_c\equiv 1/\beta_c$ in the region $1\ge p \ge p_N$.

The estimates of $T_c$ shown in Fig.~\ref{bcfig} hint at a smooth linear
behavior for small values of $w\equiv 1-p$, close to the Ising point at $w=0$.
This can be explained by some considerations on the multicritical behavior
around the Ising point at $w=0$. The Ising critical behavior at $w=0$ is
unstable against the RG perturbation induced by quenched disorder at $w>0$,
\cite{Harris-74} which leads to the RDIs critical behavior.  Indeed such a
perturbation has a positive RG dimension $y_w$ at the Ising fixed point:
\cite{Aharony-76,PV-02} $y_w=\alpha_{\rm Is}/\nu_{\rm Is}=2/\nu_{\rm Is}-3$
where $\alpha_{\rm Is}$ and $\nu_{\rm Is}$ are the Ising specific-heat and
correlation-length critical exponents, and therefore~\cite{CPRV-02}
$y_w=0.1740(8)$.  Thus, in the absence of an external magnetic field,
beside the scaling field $u_t$ related to the temperature, there is another
relevant scaling field $u_w$ associated with the quenched disorder parameter
$w\equiv 1-p$.  General RG scaling arguments~\cite{LS-84,CPPV-04} show that
the singular part of the free energy for $w \to 0$ behaves as
\begin{eqnarray}
{\cal F}_{\rm sing} \sim u_t^{2-\alpha_{\rm Is}} F(X), \qquad
X=u_w u_t^{-\phi},
\label{scalmc} 
\end{eqnarray}
where $\phi=y_w\nu_{\rm Is}=\alpha_{\rm Is}=0.1096(5)$ is the crossover
exponent, and $F(X)$ is a crossover scaling function which is universal (apart
from normalizations).  The scaling fields $u_t$ and $u_w$ depend on the
parameters of the model.  In general, we expect
\begin{equation} 
u_t = t + k w,
\end{equation}
where $t \equiv T/T_{\rm Is} - 1$, $T_{\rm Is}$ is the critical temperature of
the Ising model, and $k$ is a constant.  No such mixing between $t$ and $w$
occurs in $u_w$, since $u_w$ vanishes for $w = 0$.  Hence, we can take $u_w =
w $.  The system has a critical transition for $w>0$ at $T_c(w)$. Since the
singular part of the free energy close to a critical point behaves as $(T -
T_c)^{2-\alpha}$ ($\alpha=-0.049(6)$ is the specific-heat exponent of the RDIs
universality class), we must have $F(X_c) = 0$, where $X_c$ is the value of
$X$ obtained by setting $T = T_c(w)$ (see, e.g., Ref.~\onlinecite{PV-07} and
references therein).  Hence, we obtain
\begin{equation}
w \left[ {T_c(w)/ T_{\rm Is}} - 1 + k w \right]^{-\phi} = X_c,
\end{equation}
and therefore
\begin{eqnarray}
T_c(w)/T_{\rm Is}  - 1  = (w/X_c)^{1/\phi} - k w + \cdots,
\label{T-beh-0}
\end{eqnarray}
where the dots indicate higher-order terms.  This expression provides the $w$
dependence of the critical temperature for $w$ small.  Note that the
nonanalytic term in Eq.~(\ref{T-beh-0}) is suppressed with respect to the
analytic ones, because $1/\phi\approx 9.1$. Thus, $T_c(w) \approx T_{\rm Is}
(1 - k w + O(w^2))$.  Since $T_c(w)<T_{\rm Is}$, we can also infer that $k>0$.
From the results for $T_c(w)$ we estimate 
$k\approx 2.2$ for the $\pm J$ Ising model.

\section{Conclusions}
\label{conclusions}

In this paper we have studied the critical behavior of the 3D $\pm J$ Ising
model at the transition line between the paramagnetic and the ferromagnetic
phase, which extends from $p=1$ to a multicritical (Nishimori) point at
$p=p_N\approx 0.767$. We presented a FSS analysis of MC simulations at
various values of $p$ in the region $p_N<p<1$. The results for the critical
exponents and other universal quantities are consistent with those of the RDIs
universality class.  For example, we obtained $\nu=0.682(3)$ and
$\eta=0.036(2)$, which are in good agreement with the presently most accurate
estimates \cite{HPPV-07} $\nu=0.683(2)$ and $\eta=0.036(1)$ for the 3D RDIs
universality class. Therefore, our FSS analysis provides a strong evidence
that the critical behavior of the 3D $\pm J$ Ising along the ferromagnetic
line belongs to the 3D RDIs universality class.

We also note that the random-exchange interaction in the $\pm J$ Ising model
gives rise to frustration, while the RDIs universality class describes
transitions in generic diluted Ising systems with ferromagnetic exchange
interactions.  This implies that frustration is irrelevant at the
ferromagnetic transition line of the 3D $\pm J$ Ising model.  Moreover, the
observed scaling corrections are consistent with the RDIs leading and
next-to-leading scaling correction exponents $\omega=0.33(3)$ and
$\omega_2=0.82(8)$.  This indicates that frustration does not introduce new
irrelevant perturbations at the RDIs fixed point with RG dimension $y_f\gtrsim
-1$.

\appendix

\section{Notations}
\label{notations}

We define the two-point correlation function
\begin{eqnarray}
G(x) \equiv \overline{\langle \sigma_0 \,\sigma_x \rangle} ,
\label{twop}
\end{eqnarray}
where the overline indicates the quenched average over the $J_{xy}$
probability distribution. Then, we define the corresponding susceptibility
$\chi\equiv \sum_x G(x)$ and the correlation length $\xi$
\begin{equation}
\xi^2 \equiv {\widetilde{G}(0) - \widetilde{G}(q_{\rm min}) \over 
          \hat{q}_{\rm min}^2 \widetilde{G}(q_{\rm min}) },
\end{equation}
where $q_{\rm min} \equiv (2\pi/L,0,0)$, $\hat{q} \equiv 2 \sin q/2$, and
$\widetilde{G}(q)$ is the Fourier transform of $G(x)$.  We also consider
quantities that are invariant under RG transformations in the critical limit.
Beside the ratio
\begin{equation}
R_\xi \equiv \xi/L,
\label{rxi}
\end{equation}
we consider the quartic cumulants $U_4$, $U_{22}$ and $U_d$ defined by
\begin{eqnarray}
&& U_{4}  \equiv { \overline{\mu_4}\over \overline{\mu_2}^{2}}, 
\label{cumulants}\\
&&U_{22} \equiv  {\overline{\mu_2^2}-\overline{\mu_2}^2 \over \overline{\mu_2}^2},
\nonumber \\
&&U_d \equiv U_4 - U_{22},
\nonumber
\end{eqnarray}
where
\begin{eqnarray}
\mu_{k} \equiv \langle \; ( \sum_x \sigma_x\; )^k \rangle \; .
\end{eqnarray}
We also define corresponding quantities
$\bar{U}_4$,  $\bar{U}_{22}$,  and $\bar{U}_d$ 
at fixed $R_\xi=0.5943$.  
Finally, we consider the derivative $R'_\xi$ of $R_\xi$, 
and $U'_4$ of $U_4$,
with respect to $\beta\equiv 1/T$,
which allow one to determine the critical exponent $\nu$.


\begin{thebibliography}{99}

\bibitem{Nishimori-81}
H. Nishimori, Prog. Theor. Phys. {\bf 66}, 1169 (1981).

\bibitem{LB-88}
P. Le Doussal and A.B. Harris,
Phys. Rev. Lett. {\bf 61}, 625 (1988).

\bibitem{KR-03}
N. Kawashima and H. Rieger,  in 
{\em Frustrated Spin Systems}, edited by H.T. Diep
(World Scientific, Singapore, 2004); cond-mat/0312432.

\bibitem{KKY-06}
H. Katzgraber, M. K\"orner, and A.P. Young,
Phys. Rev. B {\bf 73}, 224432 (2006).

\bibitem{ON-87}
Y. Ozeki and H. Nishimori, 
J. Phys. Soc. Japan {\bf 56}, 3265 (1987).

\bibitem{Singh-91}
R.R.P. Singh, 
Phys. Rev. Lett. {\bf 67}, 899 (1991).

\bibitem{OI-98}
Y. Ozeki and N. Ito, J. Phys. A {\bf 31}, 5451 (1998).

\bibitem{pnest} 
Refs.~\onlinecite{ON-87,Singh-91,OI-98} report the estimates 
$p_N=0.767(2)$, $p_N=0.7656(20)$, and
$p_N=0.7673(3)$ respectively.


\bibitem{Hukushima-00}
K. Hukushima, J. Phys. Soc. Japan {\bf 69}, 631 (2000).

\bibitem{PV-02}
A. Pelissetto and E. Vicari,
Phys. Rept. {\bf 368}, 549 (2002).

\bibitem{FHY-03}
R. Folk, Yu. Holovatch, and T. Yavors'kii,
Uspekhi Fiz. Nauk {\bf 173}, 175 (2003)
[Phys. Usp. {\bf 46}, 175 (2003)].

\bibitem{BFMMPR-99}
H.~G.~Ballesteros, L.~A.~Fern\'andez, V.~Mart\'{\i}n-Mayor, A.~Mu\~noz~Sudupe,
G.~Parisi, and J.~J.~Ruiz-Lorenzo,
J. Phys. A {\bf 32}, 1 (1999).

\bibitem{HPPV-07}
M. Hasenbusch, F. Parisen Toldin, A. Pelissetto, and E. Vicari,
J. Stat. Mech.: Theory Exp. P02016 (2007). 


\bibitem{CMPV-03}
P. Calabrese, V. Mart\'{\i}n-Mayor, A. Pelissetto, and E. Vicari,
Phys. Rev. E  {\bf 68}, 036136 (2003).


\bibitem{PV-00}
A. Pelissetto and E. Vicari,
Phys. Rev. B {\bf 62}, 6393  (2000).

\bibitem{BFMMPR-98}
H.G. Ballesteros, L.A. Fern\'andez, V. Mart\'{\i}n-Mayor, A. Mu\~noz~Sudupe,
G. Parisi, and J.J. Ruiz-Lorenzo,
Phys. Rev. B {\bf 58}, 2740 (1998).

\bibitem{CPRV-02}
M. Campostrini, A. Pelissetto, P. Rossi, and E. Vicari,
Phys. Rev. E {\bf 65}, 066127 (2002).

\bibitem{DB-03}
Y. Deng  and  H.W.J. Bl\"ote,
Phys. Rev. E {\bf 68}, 036125 (2003).

\bibitem{Janke-POS} 
W. Janke, in Proceedings of the XXIII International Symposium on
Lattice Field Theory, Dublin, July 2005, 
POS(LAT2005)018

\bibitem{IOK-99}
N. Ito, Y. Ozeki, and H. Kitatani,
J. Phys. Soc. Jpn. {\bf 68}, 803 (1999).

\bibitem{CPPV-04}
P. Calabrese, P. Parruccini, A. Pelissetto, and E. Vicari,
Phys. Rev. E {\bf 69}, 036120 (2004).

\bibitem{Has-99}
M. Hasenbusch,
J. Phys. A  {\bf 32},  4851 (1999).

\bibitem{CHPV-06}
M.  Campostrini, M. Hasenbusch, A. Pelissetto, and E. Vicari,
Phys. Rev. B {\bf 74}, 144506 (2006);
Phys. Rev. B {\bf 63}, 214503 (2001).


\bibitem{multispin} 
See, e.g., S. Wansleben, J.B. Zabolitzky, and C. Kalle, J.  Stat. Phys.  {\bf
37}, 271 (1984); G. Bhanot, D. Duke, and R. Salvador, 
Phys. Rev. B {\bf 33}, 7841 (1986).

\bibitem{ranlxd}
M. L\"uscher,  Comput. Phys. Commun. {\bf 79}, 100 (1994).

\bibitem{twister}
The SIMD-oriented fast Marsenne twister random number generator 
has been introduced by
M. Matsumoto and M. Saito. Details can be found in 
M. Saito, Master Thesis (2007) and at 
{\tt http://www.math.sci.hiroshima-u.ac.jp/~m-mat/MT/emt.html}.

\bibitem{footnote1} 
  In order to make the use of these expensive (in terms of CPU-time)
  generators affordable, we employed the same sequence of random numbers for
  the update of all $n_{\rm bit}$ systems (for the initialization of
  the configurations at the beginning of the simulation we used
  independent random numbers for each of the systems).  This may give rise to
  a statistical correlation among the $n_{\rm bit}$ systems.  
  This effect is probably small and we have not detected it.  
  Anyway, in order to ensure a correct estimate
  of the statistical error, all $n_{\rm bit}$ systems that use the same sequence
  of random numbers have been put in the same bin in our jackknife analysis.

\bibitem{SW-87}
R.H. Swendsen and  J-S. Wang, 
Phys. Rev. Lett.  {\bf 58},  86 (1987).

\bibitem{Wolff-89}
U. Wolff, Phys. Rev. Lett. 
{\bf 62}, 361 (1989).

\bibitem{IIBH-06}
D. Ivaneyko, J. Ilnytskyi, B. Berche, and Yu. Holovatch,
Physica A {\bf 370}, 163 (2006). 

\bibitem{BFMMPR-98-b}
H.G. Ballesteros, L.A. Fern\'andez, V. Mart\'{\i}n-Mayor, A. Mu\~noz~Sudupe,
G. Parisi, and J.J. Ruiz-Lorenzo,
Nucl. Phys. B {\bf 512}, 681 (1998).

\bibitem{footnotecrossover} 
   The crossover exponent from pure Ising to RDIs critical behavior is the
  Ising specific-heat exponent~\cite{Aharony-76} $\alpha_{\rm Is}$, see also
  Sec.~\ref{crittc}.  This implies that the crossover scaling variable in the
  FSS at $T_c$ is given by the combination $X= c w L^{\alpha_{\rm Is}/\nu_{\rm
      Is}}$, where $w=1-p$, $\alpha_{\rm Is}/\nu_{\rm Is}=0.1740(8)$, and $c$
  is a normalization constant. When $w\to 0$, strong crossover effects are
  expected for $X\lesssim 1$, which corresponds to $L \lesssim (cw)^{-5.75}$.
  The RDIs asymptotic critical behavior is observed for $X\gg 1$.

\bibitem{Harris-74}
A.~B.~Harris, J. Phys. C {\bf 7}, 1671 (1974).

\bibitem{Aharony-76} 
A.~Aharony, in
{\em Phase Transitions and Critical Phenomena},
edited by C.~Domb and J.~Lebowitz 
(Academic Press, New York, 1976),
Vol.\ 6, p. 357.


\bibitem{LS-84}
I.D. Lawrie and S. Sarbach, in {\em Phase Transitions and Critical Phenomena},
Vol. 9, edited by C. Domb and J. Lebowitz (Academic, London, 1984).

\bibitem{PV-07}
A. Pelissetto and E. Vicari,
cond-mat/0702273.


\end{thebibliography}
\end{document}